# Probing defect induced room temperature ferromagnetism in CVD grown MoO$_3$ flakes: A correlation with electronic structure and first principle-based calculations

*Sharmistha Dey[a], Vikash Mishra[b], Neetesh Dhakar[c], Sunil Kumar[c], Pankaj Srivastava[a], and Santanu Ghosh[a*]*

[a] Nanostech Laboratory, Department of Physics, Indian Institute of Technology Delhi, New Delhi 110016, India

[b] Department of Physics, Nano Functional Materials Technology Centre and Materials Science Research Centre, Indian Institute of Technology Madras, Chennai, 600036, India

[c] Department of Physics, Indian Institute of Technology Delhi, New Delhi 110016, India

**Abstract:**

In this paper, we report the growth of pure $\alpha$-MoO$_3$ micro-flakes by CVD technique and their structural, electronic, optical, and magnetic properties. Samples are annealed at various temperatures in an H$_2$ atmosphere to induce ferromagnetism. All the samples exhibit ferromagnetism at room temperature, and 250°C annealed sample shows the highest magnetic moment of 0.087 emu/g. It is evident from PL data that pristine as well as annealed samples contain different types of defects like oxygen vacancies, surface defects, interstitial oxygen, etc. It is deduced from the analysis of Mo3d and O1s core-level XPS spectra that oxygen vacancies increase up to an annealing temperature of 250°C that correlates with the magnetic moment. Significant changes in the total density of states and also in the magnetic moment for two and three oxygen vacancies are noticed through first-principle-based calculations. It is concluded that the magnetic moment is produced by oxygen vacancies or vacancy clusters, which is consistent with our experimental findings.

* Corresponding author: santanu1@physics.iitd.ac.in

**Introduction:**

Electrical charge and spin are both employed simultaneously in spintronic devices. To employ an electrical device as a memory device, charge and spin should play an important role at the same time. For this reason, the use of diluted magnetic semiconductors (DMS) in memory devices is crucial [1]. To develop room-temperature ferromagnetism in non-magnetic semiconductors, researchers begin to dope magnetic materials, but there is always a solubility problem due to different crystal structures [2]. Defects are omnipresent in solids. A significant number of defects in solids can alter their electrical, optical, and magnetic properties [3]. Spintronics applications will greatly benefit if any non-magnetic semiconductor behaves like a magnetic semiconductor as a result of defects since there won't be a solubility issue. It raises a new topic of research in spintronic applications after revealing room-temperature ferromagnetism (RTFM) in the non-magnetic oxide $HfO_2$ [4]. After that, there are a lot of reports of RTFM in nonmagnetic oxides, such as ZnO [5], MgO [6], $SnO_2$ [3], etc. In recent days, among transition metal oxides, Molybdenum trioxide ($MoO_3$) has attracted tremendous attention due to its unique properties such as wide band gap (3.2 eV) [7], layered structure [8], heterogeneous catalysis property [9], gas sensing property [10], etc. $MoO_3$ crystallizes in layered structures with strong in-plane covalent bonds and weak van der Waals interactions between layers [11]. Nevertheless, only a small number of research papers have been published recently on ferromagnetism in $MoO_3$. There are some papers where transition metal doping-induced magnetism in $MoO_3$ [1,12–14] and oxygen vacancy-induced ferromagnetism in $MoO_3$ nanostructures [15,16] are reported. The focus of this study is ferromagnetism in $MoO_3$ caused by hydrogen annealing. $MoO_3$ micro-flakes are deposited using Chemical Vapor Deposition (CVD) method and then annealed in an environment of hydrogen at different temperatures for 2 hours. Hydrogen annealing is an efficient method for modifying the electronic and magnetic characteristics of $MoO_3$ micro-flakes. The electronic structure as well as magnetic and optical properties of pristine and annealed samples were studied. The origin of room-temperature ferromagnetism (RTFM) is elaborately elucidated and correlated to the oxygen vacancies that are induced due to annealing. Density functional theory (DFT) was carried out to explain the experimental findings of this work. A good correlation between RTFM, electronic structure, and DFT-based calculations is established.

**Experimental and computational details:**

**Experimental:**

$MoO_3$ micro-flakes were deposited by the typical CVD process on a $SiO_2$/Si substrate using a microprocessor-programmable single-zone tubular furnace. The schematic diagram is shown in Figure 1. $MoO_3$ nanopowder (99.9% pure, particle size 30–50 nm, 21 mg) was used as a precursor. $MoO_3$ powder was kept at a temperature of 850°C (central zone of the furnace) in a quartz boat, and substrate was kept at 550°C (13 cm from the central zone). A flow of 30 sccm Ar gas was used as a carrier gas. To obtain the flakes, the substrate position (distance from the central zone and also the substrate height from the base) is important. A boat of 15mm height was taken to keep the substrate such as the substrate will be in the dense gas flow region. The furnace was heated at a 12°C /minute ramp rate to reach 850°C, the deposition time was kept at 20 minutes, and then the heater was turned off for natural cooling.

The samples were annealed thereafter at three different temperatures (150°C, 250°C, and 350°C) at 50 sccm $H_2$ gas flow for 2 hours. The schematic diagram is shown in Figure 2.

For the identification of the crystalline phase, X-ray diffraction (XRD) was performed. The XRD pattern was taken in typical grazing incidence XRD mode (≤1°) by PANalytical X'Pert³ in 2θ range from 20°-60°. Raman spectroscopy was done by Renishaw Micro Raman Spectroscope. A diode laser with a 532nm wavelength was used as a source. Field emission scanning electron microscopy (FESEM) was done by secondary electron mode in TESCAN instruments and an Electron probe micro-analyzer (EPMA) was done by SHIMADZU EPMA-1720. Photoluminescence (PL) spectra were recorded by using a 325nm excitation wavelength from a He-Cd laser and HORIBA iHR550 spectrometer in side scattering geometry. X-ray photoelectron spectroscopy (XPS) spectra were recorded using AXIS Supra with the monochromatic Al Kα X-ray source (1486.6 eV). The pass energy used was 20 eV and the overall resolution was ~ 0.6 eV. The pressure during the measurements was ~$3\times10^{-9}$ mbar. Magnetic measurements were done by MPMS3 device by Quantum Design at room temperature. UV-Visible is done by Perkin Elmer instruments at room temperature.

Hereafter, the pristine, 150°C, 250°C and 350°C annealed samples will be denoted by M1, M2, M3, and M4 respectively.

**Computational details:**

Projected-augmented wave (PAW) potentials, included in the Quantum Espresso simulation package [17], have been used to conduct spin-polarized calculations on the basis of density-functional theory (DFT) [18,19]. To account for electronic correlation effects (i.e., $U_{eff}$) the Perdew-Burke-Ernzerhof (PBE) of the generalised gradient approximation has been employed with an onsite coulomb repulsion [20–22]. $U_{eff}$ = 4 eV for the Mo-d orbital and a plane wave cut-off energy of 550 eV for pure $MoO_3$ were used to calculate the band gap, which was determined to be 3.21 eV (experimental bandgap 3.2 eV). The orthorhombic structure of $MoO_3$ has been considered with 16 atoms in a unit cell and $3 \times 2 \times 2$ super cell (192 atoms) of $MoO_3$. The calculations have been made after achieving energy convergence of $5 \times 10^{-5}$ eV and the forces per atom are declined to less than 0.04 eV/A° for pure, defective, and oxygen vacancy incorporated $MoO_3$.

**Results and discussion:**

Figure 3 shows the XRD patterns recorded at room temperature for pristine samples and also for $H_2$-annealed samples. From the XRD patterns, we confirm the formation of pure α- $MoO_3$ phases which has an orthorhombic lattice structure with space group *Pbnm* [23]. As the flakes are randomly oriented on the substrate, the intensity of the particular peaks varied with different annealing temperatures. All peaks are identified with appropriate (hkl) values, which are clearly denoted in Figure 3 [24]. We can clearly see the change in intensity of the peaks after annealing, which confirms structural changes in the samples.

Figure 4 shows the Raman spectra of as-grown and annealed samples at room temperature. From the Raman spectrum also, we can conclude the formation of pure α- $MoO_3$ phases. All peaks are identified with appropriate modes of vibration and denoted clearly in Figure 4, which arise due to the Mo-O bending and stretching vibrations, respectively [25]. The peak at 996 cm$^{-1}$ is due to the terminal oxygen ($Mo^{6+}$=O) stretching mode, which is caused by unshared oxygen [26]. The peak at 817 cm$^{-1}$ is due to the doubly coordinated oxygen ($Mo_2$-O) stretching mode, which is caused by corner-sharing oxygen atoms common to two octahedra. The peak at 664 cm$^{-1}$ is due to the triply coordinated oxygen ($Mo_3$-O) stretching mode, which is caused by edge-sharing oxygen atoms common to three octahedra. The 336 and 377 cm$^{-1}$ peaks are due to O-Mo-O bending modes. The 283 cm$^{-1}$ peak is due to the wagging mode for the double bond O=Mo=O. The 197, 215, and 242 cm$^{-1}$ peaks are due to O=Mo=O twist modes. The 157 and 291 cm$^{-1}$ peaks are due to the O=Mo=O wagging modes. The 469 cm$^{-1}$ peaks correspond to the O-Mo-O stretching mode [27]. The asterisk peak is identified as the Si substrate peak.

Figure 5 shows the FESEM images of pristine and H$_2$-annealed samples. As shown in Figure 5(a), the pure MoO$_3$ samples show a micro-belt-like structure, and the average thickness is in the order of 2 μm (which is also confirmed from cross-sectional FESEM and is shown in the supplementary information (Fig.S1)). After annealing in the H$_2$ atmosphere, morphological changes have occurred, which are shown clearly in Figures 5 (b), (c), and (d). The breadth of the samples decreased with increasing annealing temperature, and it became very narrow for M4 samples. So, it is obvious that the material is degraded for M4 samples.

EPMA is used for mapping and quantitative and qualitative analysis of as-grown and 250°C annealed samples, as we get the highest magnetic moment for 250°C annealed samples (discussed later). Only Molybdenum (Mo) and Oxygen (O) are present in both samples, confirming their purity. From the Pristine to M3 samples, the surface alteration is clearly visible in the supplementary information (Fig.S2)). It is evident that the pristine sample has a comparatively smooth surface, whereas the annealed sample has a rough surface. The mapping and qualitative analysis for the pristine sample are shown in Figure 6. At first, we selected an area, which is shown in Figure 6(a), and carried out mapping independently for Mo and O elements. For Mo, we use a PET, and for O, we use a RAP crystal detector. Figure 6(b) shows the mapping for O, and Figure 6(c) shows the mapping for Mo. Figure 6(d) and (e) show the qualitative analysis for Mo and O. Au signal is coming from the coating which is used to avoid the charging effect and Si is coming from the substrate. From the quantitative analysis, we can clearly see that there is a formation of oxygen vacancies in annealed samples, which is shown in Table 1 and Table 2 and supports our SQUID results. The mapping and qualitative analysis for the 250°C annealed samples are shown in the supplementary information ( Fig.S3).

To comprehend the significance of various types of defects present in our sample photoluminescence (PL) spectroscopy is carried out at room temperature. Figure 7 shows the PL spectra of as-grown and H$_2$-annealed MoO$_3$ samples by using a 325 nm excitation wavelength. The wide peak of the PL profile has a wavelength range of 400 to 530 nm and is dominated by the blue and green luminescence bands. Generally, in MoO$_3$, there are different kinds of intrinsic defects such as Mo$_i$, O$_i$, V$_{Mo}$, O$_{Mo}$, V$_{Mo}^-$, V$_o$, and V$_o^+$, etc. We deconvoluted the peaks using seven Gaussian peaks while keeping the FWHM constant for all the corresponding peaks. The peaks are located at ~2.35eV, ~2.43eV, ~2.56eV, ~2.73eV, ~2.84eV,

~2.98eV and ~3.19 eV. The peaks at ~2.98eV and ~3.19 eV represent the band-to-band (CB to VB) transition [14]. The green emission peak at ~2.35eV and ~2.43eV is due to the recombination between the CB and VB of $MoO_3$ micro-flakes [28]. There is also a different explanation behind the origin of these peaks, such as the presence of Mo interstitial sites, surface defects, or $Mo_xO_y$ variations. The peak located at ~2.56 eV is for various h-$MoO_3$ hexagonal phases. The blue emission peak located at ~2.73eV is due to oxygen vacancy-type defects [15]. The visible emissions may have been caused by transitions of excited optical centers in the deep levels (DL), which are often caused by structural and surface defects [29]. From PL spectra, we can conclude that CVD-grown M1 sample and the annealed samples contain different types of defects.

To investigate the electronic structure, XPS is performed at room temperature. The Mo 3d core level XPS spectra are shown in Figure 8. The XPS spectrum is deconvoluted into Gaussian peaks. As evident, $Mo^{6+}$ valence states predominate in all the samples, which confirms pure $MoO_3$ phase formation. $3d_{5/2}$ and $3d_{3/2}$ are located at 232.8 and 236 eV, respectively. Additionally, the presence of $Mo^{5+}$ is less as compared to $Mo^{6+}$. Up to M3 samples, the area beneath the $Mo^{5+}$ state increases with annealing temperature, and for M4 samples, it decreases. The production of the $Mo^{4+}$ states $3d_{5/2}$ and $3d_{3/2}$ in M4 samples occurs at 229.5 and 232.5 eV, respectively. As $MoO_2$ is diamagnetic in nature, it shows less ferromagnetism than M3 samples [2].

The O 1s core level spectra are shown in Figure 9, from which it is evident that new peaks are produced when the annealing temperature is raised. For the Pristine sample, the O1s XPS spectra are deconvoluted into three Gaussian peaks. The peak at ~530.6 eV is ascribed to the lattice oxygen, i.e., Mo bonded with lattice oxygen. The peak at ~531.5 eV corresponds to oxygen vacancies [16]. The peak at ~533 eV is due to the chemically absorbed oxygen at the surface. The peak intensity of the oxygen vacancy-related peak is increasing up to M3 samples, and the peak intensity of the absorbed oxygen peak is decreasing gradually which satisfies our SQUID results. The peaks at ~532.7 eV, ~533.4 eV, ~534 eV, and ~535.2 eV for the high-temperature annealed samples are due to the additional hydroxyl group, $H_2O$ or Mo-OH bonds, etc [2,30]. The peaks at ~534 and ~535.2eV in O1s spectra are due to C-OH [31] and adsorbed oxygen, respectively [32].

The diffuse reflectance spectra (DRS) for all the samples at room temperature in the 200–800 nm wavelength range is shown in the supplementary information (Fig.S4). All the samples have very low reflectance, i.e., within 15%. We got the lowest reflectance for the M3 sample because it contains more oxygen vacancies than other samples. Through XPS and EPMA analysis, we can observe that lattice defects, vacancies, and surface defects are growing as a result of annealing. Due to the increasing number of defects, absorption is increasing, or the accumulation of defects that act as colour centres is the cause of the reduction of reflectance [25].

The M-H loops as obtained by SQUID measurements for pristine and the $H_2$ annealed samples at room temperature are shown in Figure 10 (a). It is clear from these plots that all samples exhibit ferromagnetism at room temperature. The magnetic moment gradually increases with annealing temperature, and then it decreases at the highest annealing temperature. Saturation magnetization, coercivity, and remnant magnetization values of all samples are shown in table 3. By XPS and EPMA analysis, it is evident that the oxygen vacancy increases with increasing annealing temperature; this indicates the magnetic moment is also increasing. Also, no magnetic impurities like Fe, Ni, Co, etc. in the pristine and as well as in the annealed samples are observed from EPMA analysis which clearly indicates that FM in the samples are intrinsically originated. From Figure 10 (b), we can clearly see that there is a one-to-one correspondence between the magnetic moment and oxygen vacancies. There are some reports that mention oxygen vacancy-induced ferromagnetism in nonmagnetic oxides at ambient temperature [33–35]. In addition to changing the band structure of the host oxides, the vacancies have the ability to modify the valence of nearby elements. Both of these effects have the potential to significantly contribute to ferromagnetism. An extended bound magnetic polarons (BMP) model called the F center exchange mechanism was also put forth to explain the $d^0$ ferromagnetism, which is caused by oxygen vacancies. Through this process, the oxygen vacancies on the surfaces of the α-MoO$_3$ are the origin of the unpaired electron spins responsible for ferromagnetism. It is reasonable to assume that the polarised electrons in the singly occupied oxygen vacancies embedded deep in the gap will result in RTFM [15]. Also, Patel *et al.* assume that a significant factor in modifying the long-range ferromagnetic ordering is the unpaired electron spins in the oxygen vacancies, which are primarily found near the surface [36]. Hence, the rise in saturation magnetization for $H_2$ annealed α-MoO3 micro-flakes might be due to the enhanced exchange interactions between localised electron spins caused by oxygen vacancies. Although the exact characteristics of ferromagnetic exchange interactions are still unclear.

Additionally, first-principles-based calculations were performed to understand the ferromagnetic behavior of O-vacancy-incorporated $MoO_3$ and its inherent defects. $MoO_3$ structures with different oxygen vacancy configurations have been shown in Figure 11. First, we started with the electronic properties of pure $MoO_3$ in a $3 \times 2 \times 2$ supercell. Figure 12 (a) displays the total density of states (TDOS) of pure stoichiometric $MoO_3$ (i.e., the total density of states for spin up and spin down are the same), and it suggests the non-magnetic behavior of pure $MoO_3$. Following up on our experimental findings, we ran calculations using oxygen vacancies. First, we introduce single oxygen vacancy (Figure 12 (b)), and it is evident that there is a small change in TDOS and also the appearance of a magnetic moment (Table 4). Since oxygen vacancies are the primary cause of ferromagnetism according to our experimental findings, calculations are performed with various combinations of oxygen vacancies. After that, two and three oxygen vacancies are introduced in the supercell for calculations. Remarkable changes in TDOS and also in the magnetic moment for two and three oxygen vacancies are noticed (Table 4). Although the magnetic moment for Molybdenum vacancies is more than oxygen vacancies but the formation energy is comparatively higher than oxygen-type vacancies (Table 4), so the formation of Mo-type vacancies is less favorable. The 4d orbitals of the Mo atoms and the 2p orbitals of the O atoms are the primary sources of induced ferromagnetism [11]. Finally, we can draw the conclusion that the magnetic moment is produced by oxygen vacancies or vacancy clusters, which is consistent with our experimental findings.

**Conclusions:**

The $\alpha$-$MoO_3$ micro-flakes prepared by the CVD process were annealed at three different temperatures in a $H_2$ environment. For structural analysis, XRD and Raman spectroscopy are done, which confirm pure $\alpha$-$MoO_3$ formation. FESEM images show clear morphological changes after annealing. For elemental analysis and a purity check, EPMA is carried out, and it confirms only the presence of Mo and O. To know about the defects, present in the pristine as well as annealed samples, PL is performed at room temperature, and it shows the presence of different kinds of defects like oxygen vacancies, surface defects, oxygen interstitials, etc. To know about the surface composition, XPS was done, and it showed the increment of oxygen vacancies up to M3 samples, which further satisfied our SQUID and UV-Visible data. The M3 sample exhibits the highest magnetic moment and the lowest reflectance because it contains more oxygen vacancies.

**Acknowledgements:**

The authors are grateful to the Physics Department at IIT Delhi for the XRD, Raman, and SQUID facilities; the Central Research Facilities (CRF) for the FESEM, EPMA, and XPS facilities; and the Nanoscale Research Facility (NRF) for the UV-Vis spectrophotometer facility. One of the authors, Sharmistha Dey, acknowledges IIT Delhi for a Senior Research Fellowship.

**List of figures:**

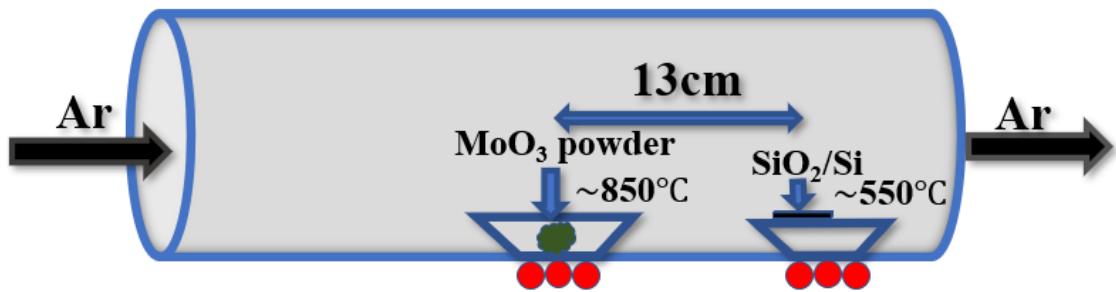

**Fig.1.** Schematic diagram for the CVD setup for the deposition of $MoO_3$.

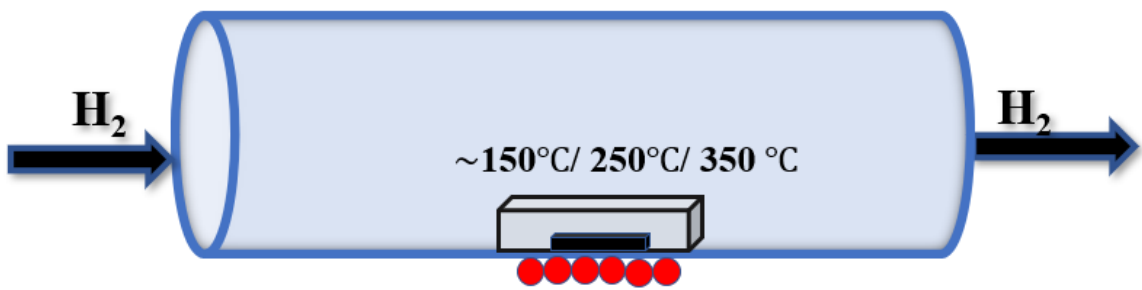

**Fig.2.** Schematic diagram for the lab setup for $H_2$ annealing of $MoO_3$.

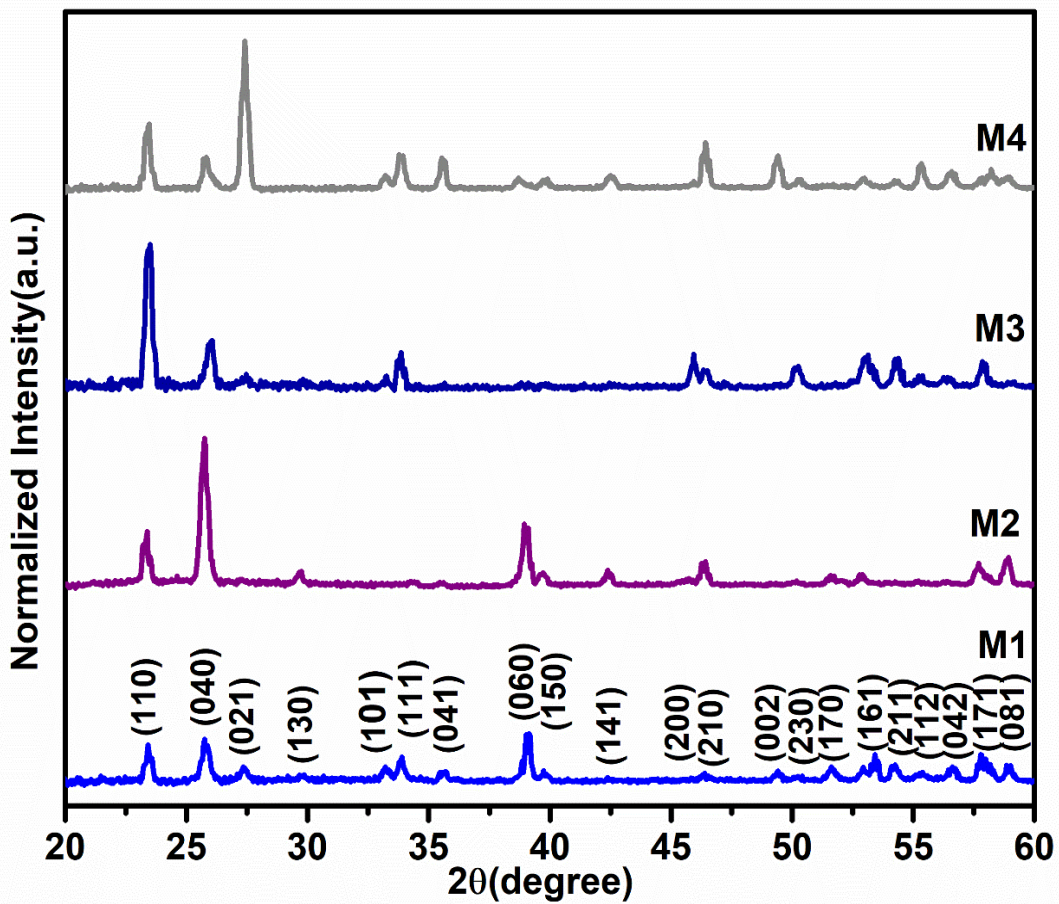

**Fig.3.** XRD pattern of Pristine and annealed $MoO_3$ micro-flakes.

12